\documentclass[pra,twocolumn,superscriptaddress]{revtex4-1}

\usepackage{amsmath}
\usepackage{latexsym}
\usepackage{amssymb}
\usepackage{graphicx}
\usepackage[colorlinks=true, citecolor=blue, urlcolor=blue]{hyperref}
\usepackage{float}
\usepackage{textcomp}
\usepackage{mathpazo}
\usepackage{bbm}

\newcommand{\tr}{{\rm Tr}}

\newcommand{\idol}{\ensuremath{\mathbbm 1}}

\newcommand{\ket}[1]{|#1\rangle}
\newcommand{\bra}[1]{\langle #1|}

\makeatletter
\usepackage{pifont}
\makeatother

\usepackage{dcolumn}
\usepackage{epstopdf}
\usepackage{subfigure}

\begin{document}

\title{Evolution equation for quantum coherence}
\author{Xinzhi Zhao}
\affiliation{School of Physical Science and Technology, Ningbo University, Ningbo, 315211, China}
\author{Jianwei Shao}
\affiliation{School of Physical Science and Technology, Ningbo University, Ningbo, 315211, China}
\author{Yi Zheng}
\affiliation{CAS Key Laboratory of Quantum Information, University of Science and Technology of China, Hefei 230026, China}
\affiliation{CAS Center for Excellence in Quantum Information and Quantum Physics, University of Science and Technology of China, Hefei 230026, China}
\author{Wen-Zhao Zhang}
\email{zhangwenzhao@nbu.edu.cn}
\affiliation{School of Physical Science and Technology, Ningbo University, Ningbo, 315211, China}
\author{Chengjie Zhang}
\email{chengjie.zhang@gmail.com}
\affiliation{School of Physical Science and Technology, Ningbo University, Ningbo, 315211, China}

\begin{abstract}
Quantum coherence plays an important role in quantum resource theory, which is strongly related with entanglement. Similar to the entanglement evolution equation, we find the coherence evolution equation of quantum states through fully and strictly incoherent operation (FSIO) channels. In order to quantify the full coherence of qudit states, we define G-coherence and convex roof of G-coherence, and prove that the G-coherence is a strong coherence monotone  and the convex roof of G-coherence is  a coherence measure under FSIO, respectively. Furthermore, we prove a coherence evolution equation for arbitrary $d$-dimensional quantum pure and mixed states under FSIO channels, which generalizes the entanglement evolution equation for bipartite pure states. Our results will play an important role in the simplification of dynamical coherence measure.
\end{abstract}
%\date{\today}

\maketitle

\section{Introduction}
Coherence and entanglement are two key quantum resources that have many similarities and applications in quantum information theory \cite{Nielsen}. To implement entanglement-based quantum protocols, it is essential to understand how entanglement evolves when one of the subsystems is subjected to a noisy channel. In Ref. \cite{Konrad}, the authors derived a simple relation that captures the dynamics of entanglement for two-qubit systems under arbitrary local channels, and showed that the entanglement can be factorized into a product of local and global contributions.

The entanglement evolution equation proposed in Ref. \cite{Konrad} has been experimentally tested by Farias \textit{et al.} using a linear optics setup, for two different initial states: a quasipure state and a highly mixed state \cite{JF}. Furthermore,  in Ref.~\cite{xu}, Xu \textit{et al.} investigated the bipartite entanglement dynamics under one-sided open system \cite{T,F,A}. They measured the entanglement of both pure and mixed two-photon states under the effects of phase damping and amplitude decay on one of the subsystems. The results in Ref. \cite{Konrad} can be generalized to two-party quantum systems with arbitrary finite dimensions \cite{Tiersch}. Furthermore, a evolution equation for multipartite entanglement \cite{Carvalho} was derived in Ref. \cite{Gour}, which describes how the entanglement of a composite quantum system changes when one of the subsystems undergoes a physical process.

Quantum coherence is not only a manifestation of the superposition principle of quantum states, but also a crucial feature of quantum mechanics. It has been shown that coherence can be regarded as a physical resource \cite{Baumgratz,Streltsov,M.-L,Chitambar,Yuan,Hu,Zhu,SunLL,ChenL,LuH,LuH2,Fei,Fei2,Fei3,Zhang1,Zhang2,Zhang3,Qi,Hu2,xiang1,xiang2,xiang3,Liu1,Liu2,Liu3}, similar to entanglement \cite{ent1,ent2,ent3,ent3,ent4,ent5,ent6,ent7}. One of the main challenges of quantum coherence theory is to find effective methods to capture the dynamics of coherence. Moreover, finding a general law that determines the evolution equation of coherence could help us to design such methods. In Ref.~\cite{Hu}, the authors introduced a framework for studying the evolution equation of coherence, and proved a simple factorization relation based on this framework. They also identified the sets of quantum channels that satisfy this factorization relation. They noted that the universality of this factorization relation implies that it can be applied to many other coherence and quantum correlation measures \cite{Yao,Xi,Ma}.

In this paper, we introduce G-coherence and convex roof of G-coherence to quantify the full coherence for qudit states, and show that they satisfy strong coherence monotone and coherence measure properties under fully and strictly incoherent operation (FSIO) channels, respectively. Furthermore, we prove a coherence evolution equation for arbitrary $d$-dimensional quantum pure and mixed states under FSIO channels, which generalizes the entanglement evolution equation for bipartite pure states. It is worth noticing that our coherence evolution equation has been verified experimentally for qubits and qutrits under genuinely incoherent operations (GIOs) \cite{exp}, which are a special case of FSIO channels.

\section{Strong coherence monotone and coherence measure under FSIO}
Before introducing the definition of FSIO, let us review the concepts of  GIOs and fully incoherent operations (FIOs) \cite{de}.
\begin{figure}
    \centering
    \includegraphics[scale=0.5]{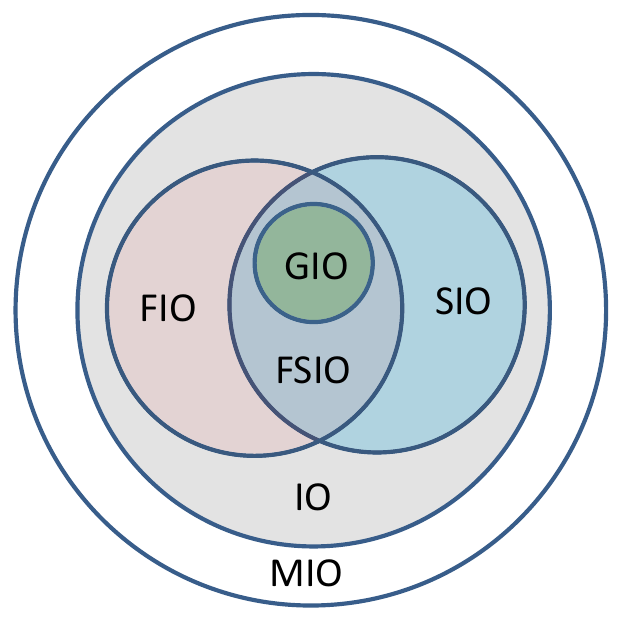}
    \caption{The relationship among GIO, FSIO, FIO, SIO, IO, and MIO. One can see that FSIO is a subset of FIO, and also a subset of SIO,  i.e., GIO$\subset$ FSIO$\subset$ FIO $\subset$ IO $\subset$ MIO and GIO$\subset$ FSIO$\subset$ SIO $\subset$ IO $\subset$ MIO.}\label{Fig1}
\end{figure}
A quantum channel $\Phi$ can be written as a set of Kraus operators $K_n$ acting on the density matrix of a quantum state $\rho$,
\begin{equation}\label{5}
  \Phi(\rho)=\sum_n{K_n\rho K_n^\dagger},
\end{equation}
with $\sum_n{K_n^\dagger K_n}=\textit{\idol}$. As shown in Ref. \cite{de}, GIO channels have been defined that all Kraus operators $\{K_n\}$ of GIO are diagonal matrices under the reference basis, and satisfy $K_n\mathcal{I}K_n^\dagger\subset \mathcal{I}$, where $\mathcal{I}$ is the set of all incoherent states. Moreover, it has also been proven that all Kraus operators of FIO are incoherent and have the same form (see Theorem 16 in Ref. \cite{de}).

\subsection{fully and strictly incoherent operations}
Now we define the FSIO channels, which can be viewed as a generalization of GIO.

A channel is FSIO if all Kraus operators are strictly incoherent and have the same form, i.e., all Kraus operators $\{K_n\}$ of FSIO channels can be written as
\begin{equation}\label{FSIO}
  K_n=U_{\pi}A_n,
\end{equation}
where $\{A_n\}$ are diagonal matrices under the reference basis satisfying $\sum_n A_n^\dag A_n=\idol$, and $U_{\pi}=\sum_{i}|\pi(i)\rangle\langle i|$ is the permutation unitary matrix with $\{|\pi(i)\rangle\}$ being a permutation of $\{|i\rangle\}$. It is worth noticing that  $U_{\pi}$ is independent of the index $n$, i.e., all Kraus operators $\{K_n\}$ have the same form.

The difference between FSIO and FIO is that all Kraus operators of FIO have at most one non-zero entry in every column, while for FSIO there is at most one non-zero entry in  not only every column but also every row.
From Fig.~(\ref{Fig1}), one can see that FSIO is a subset of FIO, and also a subset of SIO. The relationship among GIO, FSIO, FIO, SIO, incoherent operation (IO), and maximally incoherent operation (MIO) has been shown in Fig. \ref{Fig1}, i.e., GIO$\subset$ FSIO$\subset$ FIO $\subset$ IO $\subset$ MIO and GIO$\subset$ FSIO$\subset$ SIO $\subset$ IO $\subset$ MIO. For instance,
\begin{equation}\label{}
 K_i=\begin{pmatrix}0&a_i&0\\b_i&0&0\\0&0&c_i\\
\end{pmatrix},\ \
K'_i=\begin{pmatrix}0&a_i&0\\b_i&0&c_i\\0&0&0\\
\end{pmatrix},
\end{equation}
where $\{K_i\}$ belong to FSIO, but $\{K'_i\}$ belong to FIO.

\subsection{G-coherence  and convex roof of G-coherence}
The distance-based measures are widely adopted in coherence measures \cite{Levi,Chitambar2,Winter,Yadin}. For instance, $l_p$ norm of coherence is one of the measures based on matrix norms, which is an important distance-based coherence measure.
In Ref. \cite{Baumgratz}, the authors choose $l_1$ norm to quantify coherence. For any quantum states in finite-dimensional systems, the $l_1$ norm of coherence is
\begin{equation}\label{6}
  C_{l_{1}}(\rho)=\sum_{i,j}^{i\neq j}{|\rho_{ij}|}.
\end{equation}
It is easy to understand that the $l_1$ norm of coherence can be regarded as the sum of all moduli of off-diagonal elements in the density matrix, and $l_1$ norm of coherence is zero if and only if all off-diagonal elements of the density matrix are zero.

Let us now define G-coherence for quantifying  full coherence of quantum states in $d$-dimensional systems.

The G-coherence of a qu$d$it state $\rho$ is
\begin{equation}\label{G-coherence}
  G(\rho):=d\prod_{i,j}^{i\neq j}{|\rho_{ij}|^{\frac{1}{d(d-1)}}}.
\end{equation}

From this definition, one can see that the G-coherence will be zero if and only if there exists at least one off-diagonal density matrix element  $\rho_{ij}=0$ ($i\neq j$), which means the G-coherence quantify the full coherence of a $d$-dimensional quantum state $\rho$. When $d=2$, the G-coherence is equal to the $l_1$ norm of coherence, i.e., $G(\rho)=2|\rho_{12}|=C_{l_1}(\rho)$. Furthermore, one can also define the convex roof of G-coherence.

The  convex roof of G-coherence for a qu$d$it mixed state $\rho$ is defined as
\begin{equation}\label{15}
  \widetilde{G}(\rho):=\underset{\{p_k,\ket{\psi_k}\}}{\inf}\sum_{k}p_{k}G(\ket{\psi_{k}}\bra{\psi_{k}}),
\end{equation}
where the infimum is taken over all possible pure state decompositions of $\rho=\sum_{k}{p_k\ket{\psi_{k}}\bra{\psi_{k}}}$ with $0\leq p_k\leq1$ and $\sum_{k}{p_{k}}=1$. For pure states, $G(\ket{\psi_{k}}\bra{\psi_{k}})$ is defined by Eq. (\ref{G-coherence}).

\subsection{G-coherence as strong coherence monotone}
In this subsection, we will prove that the G-coherence is a strong coherence monotone under FSIOs.  Recall that any coherence measures $C$ should satisfy \cite{Baumgratz}: \\
(C1) $C(\delta)=0$ if $\delta\in \mathcal{I}$; \\
(C2) Monotonicity. $C(\rho)\geq C(\Lambda(\rho))$ for any incoherent operation $\Lambda$; \\
(C3) Strong monotonicity. $C(\rho)\geq\sum_i q_i C(\sigma_i)$ with $q_i=\tr(K_i \rho K_i^\dag)$, $\sigma_i=K_i \rho K_i^\dag/q_i$ and $K_i$ being incoherent Kraus operators; \\
(C4) Convexity. $\sum_i p_i C(\rho_i)\geq C(\sum_i p_i \rho_i)$.

If a quantity $C$ satisfies C1 and C2 (C1 and C3), then we call it a (strong) coherence monotone; If a quantity $C$ satisfies C1-C4, then we call it a coherence measure.

Let us prove that he G-coherence (\ref{G-coherence}) is a strong coherence monotone under FSIOs, i.e., it satisfies conditions C1 and C3 under FSIOs.

(C1). Based on Eq. (\ref{G-coherence}), it is easy to prove that $G(\delta)=0$ if $\delta$ is incoherent, since all off-diagonal density matrix elements of an incoherent state are always zero. Actually, $G(\delta)=0$  means $\delta$ is fully incoherent (there exists at least one off-diagonal  matrix element being zero). %The set of fully incoherent states is a subset of $\mathcal{I}$.

(C3). Now we prove the condition C3 under FSIOs:
\begin{equation}\label{mono}
  G(\rho)\geq \sum_n q_n G\Big(\frac{K_n\rho K_n^\dag}{q_n}\Big),
\end{equation}
where $q_n=\tr(K_n\rho K_n^\dag)$ and $\{K_n\}$ are Kraus operators of FSIO channels satisfying Eq. (\ref{FSIO}). Since $U_{\pi}$ in Eq. (\ref{FSIO}) is the permutation unitary matrix  which is independent of the index $n$, it does not change the G-coherence of a quantum state, i.e.,
\begin{equation}\label{Uinvariant}
G(U_{\pi}\rho U_{\pi}^\dag)=G(\rho)
\end{equation}
holds for an arbitrary quantum state $\rho$. Therefore, the right hand side (RHS) of Eq. (\ref{mono}) can be written as
\begin{eqnarray}
  \mathrm{RHS} %&=&  \sum_n q_n G\Big(\frac{K_n\rho K_n^\dag}{q_n}\Big) \nonumber \\
   &=&  \sum_n q_n G\Big(\frac{ U_{\pi}A_n\rho A_n^\dag U_{\pi}^\dag }{q_n}\Big) \nonumber \\
   &=&  \sum_n q_n G\Big(\frac{A_n\rho A_n^\dag}{q_n}\Big) \nonumber\\
   &=&  d\sum_n q_n  \prod_{i,j}^{i\neq j} \bigg|\frac{[A_n \rho A_n^\dag]_{ij}}{q_n}  \bigg|^{\frac{1}{d(d-1)}}    \nonumber\\
   &=&   d\sum_n q_n \frac{1}{q_n}  \prod_{i,j}^{i\neq j} \big|[A_n \rho A_n^\dag]_{ij}  \big|^{\frac{1}{d(d-1)}}       \nonumber\\
   &=&   d\sum_n  \prod_{i,j}^{i\neq j} \big|a_{ii}^{(n)} \rho_{ij} (a_{jj}^{(n)})^*  \big|^{\frac{1}{d(d-1)}} \nonumber\\
   &=&   d \prod_{i,j}^{i\neq j} |\rho_{ij}|^{\frac{1}{d(d-1)}}  \Big(  \sum_n  \prod_{i,j}^{i\neq j} \big|a_{ii}^{(n)} (a_{jj}^{(n)})^*  \big|^{\frac{1}{d(d-1)}}  \Big) \nonumber\\
   &=&   G(\rho)  \sum_n \prod_{i=1}^{d} \big( \big|a_{ii}^{(n)}  \big|^2\big)^{\frac{1}{d}},
\end{eqnarray}
where all $A_n=\mathrm{diag}\{\cdots,a_{ii}^{(n)},\cdots\}$ are  diagonal matrices under the reference basis satisfying $\sum_n A_n^\dag A_n=\idol$, and the last equation holds since there are $d-1$ copies of $a_{ii}^{(n)}$  and $(a_{jj}^{(n)})^*$ in the product $\prod_{i,j}^{i\neq j}$. Thus, in order to prove Eq. (\ref{mono}), one only needs to prove
\begin{equation}%\label{mono2}
 \sum_n \prod_{i=1}^{d} \big( \big|a_{ii}^{(n)}  \big|^2\big)^{\frac{1}{d}}\leq1.
\end{equation}
By using the inequality of arithmetic and geometric means, we can obtain
\begin{eqnarray}
\sum_n \prod_{i=1}^{d} \big( \big|a_{ii}^{(n)}  \big|^2\big)^{\frac{1}{d}} \leq \sum_n \frac{\sum_i \big|a_{ii}^{(n)}  \big|^2}{d} = 1 ,
\end{eqnarray}
The last equation holds, since we trace both sides of  $\sum_n A_n^\dag A_n=\idol$, and obtain $\sum_n \sum_i \big|a_{ii}^{(n)}  \big|^2=d$. Therefore, the inequality (\ref{mono}) holds.

\subsection{Convex roof of G-coherence as coherence measure}
In this subsection, we will prove that the convex roof of G-coherence (\ref{15}) is a coherence measure under FSIOs i.e., it satisfies conditions C1-C4 under FSIOs.

(C1). Based on Eq. (\ref{15}), it is easy to prove that $\widetilde{G}(\delta)=0$ if $\delta$ is incoherent.

(C3). Now we prove the condition C3 under FSIOs:
\begin{equation}%\label{mono2}
  \widetilde{G}(\rho)\geq \sum_n q_n \widetilde{G}\Big(\frac{K_n\rho K_n^\dag}{q_n}\Big),
\end{equation}
where  $q_n=\tr(K_n\rho K_n^\dag)$ and $\{K_n\}$ are Kraus operators of FSIO channels satisfying Eq. (\ref{FSIO}). Suppose that $\{p'_k,\ket{\psi'_{k}}\}$ can reach the infimum of Eq. (\ref{15}), thus one can obtain
\begin{eqnarray}
\widetilde{G}(\rho)&:=&\underset{\{p_k,\ket{\psi_k}\}}{\inf}\sum_{k}p_{k}G(\ket{\psi_{k}}\bra{\psi_{k}})  \nonumber\\
&=&\sum_{k}p'_{k}G(\ket{\psi'_{k}}\bra{\psi'_{k}})  \nonumber\\
&\geq& \sum_{k}p'_{k}\sum_n q_n^{(k)} G\Big(\frac{K_n \ket{\psi'_{k}}\bra{\psi'_{k}} K_n^\dag}{q_n^{(k)}}\Big)\nonumber\\
&=& \sum_{k}p'_{k}\sum_n  G\Big(K_n \ket{\psi'_{k}}\bra{\psi'_{k}} K_n^\dag\Big),\label{13}
\end{eqnarray}
where we have used $q_n^{(k)}=\tr(K_n \ket{\psi'_{k}}\bra{\psi'_{k}} K_n^\dag)$, $G(s \rho)=sG(\rho)$ for an arbitrary positive real number $s$, and Eq. (\ref{mono}) for $\ket{\psi'_{k}}$. On the other hand,
\begin{eqnarray}
&&\sum_n q_n \widetilde{G}\Big(\frac{K_n\rho K_n^\dag}{q_n}\Big)\nonumber\\
&=&\sum_n q_n \widetilde{G}\Big(\frac{K_n \sum_{k}p'_k \ket{\psi'_{k}}\bra{\psi'_{k}}   K_n^\dag}{q_n}\Big) \nonumber\\
&\leq&\sum_n q_n \sum_{k}p'_k G\Big(\frac{K_n \ket{\psi'_{k}}\bra{\psi'_{k}}   K_n^\dag}{q_n}\Big)  \nonumber\\
&=& \sum_{k}p'_{k}\sum_n  G\Big(K_n \ket{\psi'_{k}}\bra{\psi'_{k}} K_n^\dag\Big) \nonumber\\
&\leq& \widetilde{G}(\rho),
\end{eqnarray}
where we have used $\rho=\sum_{k}p'_k \ket{\psi'_{k}}\bra{\psi'_{k}}$, $G(s \rho)=sG(\rho)$, and Eq.~(\ref{13}). It is worth noticing that $\{q_n^{(k)},K_n \ket{\psi'_{k}}/\sqrt{q_n^{(k)}}\}$ is one kind of decomposition of the mixed state $K_n\rho K_n^\dag/q_n$, so the average is no less than the infimum of Eq. (\ref{15}).

(C4). Eq. (\ref{15}) is defined by the convex roof, so $\widetilde{G}$ is a natural convex function satisfying  $\sum_i p_i \widetilde{G}(\rho_i)\geq \widetilde{G}(\sum_i p_i \rho_i)$.

(C2). From (C3) and (C4) one can easily obtain (C2).

\section{Coherence evolution equation under FSIO}
Consider a $d$-dimensional Hilbert space $\mathcal{H}$, an arbitrary pure state in $\mathcal{H}$ can be expressed as
\begin{equation}\label{1}
  \ket{\psi}=\sum_{i=1}^{d}a_i\ket{i}.
\end{equation}
Especially, when all  $a_i=1/\sqrt{d}$, the pure state (\ref{1}) is just the maximally coherent state,
\begin{equation}\label{2}
  \ket{\psi^+}=\sum_{i=1}^{d}\frac{1}{\sqrt{d}}\ket{i}.                 %¶¨Òå×î´óÏà¸ÉÌ¬
\end{equation}
If we have a maximally coherent state $\ket{\psi^+}$, how does it transform to an arbitrary quantum state (\ref{1})? In other words,
\begin{equation}\label{3}
  \ket{\psi}=S \ket{\psi^+}.
\end{equation}
Here $S$ is a diagonal matrix, and after some simple calculation, we can get
\begin{equation}\label{4}
  S=\sum_{i=1}^d a_i\sqrt{d}|i\rangle\langle i|=\mathrm{diag}\begin{Bmatrix}a_1\sqrt{d},a_2\sqrt{d},\cdots,a_d\sqrt{d}\end{Bmatrix}.               %S¾ØÕóµÄ¾ßÌå±í´ï
\end{equation}

Before we introduce our main result, let us first review the entanglement evolution equation in two-qudit systems \cite{Tiersch}. By using the G-concurrence $G_d$ \cite{Barnum,Fan,GG}, Ref. \cite{Tiersch} showed that the entanglement evolution of a two-qudit pure state $|\chi\rangle$ with one system undergoing an arbitrary quantum channel $\$$, satisfied the entanglement evolution equation,
\begin{equation}\label{entanglement}
  G_d[\idol\otimes\$(|\chi\rangle\langle\chi|)]=G_d(|\chi\rangle)G_d[\idol\otimes\$(|\Psi^+\rangle\langle\Psi^+|)],
\end{equation}
where $|\Psi^+\rangle=\sum_i |ii\rangle/\sqrt{d}$ is the maximally entangled two-qudit state. Then we are going to introduce several theorems for pure and mixed qudit states by using the G-coherence and the convex roof of G-coherence.

\subsection{Pure single-qudit states}
Let us first introduce our results for pure single-qudit states. The theorem of coherence evolution equation by using the G-coherence (\ref{G-coherence}) is as follows.

\textbf{Theorem 1}. If $\ket{\psi}$ is an arbitrary $d$-dimensional pure state (\ref{1}), after passing through a FSIO channel $\Phi$, its coherence satisfies the following evolution equation by using the G-coherence (\ref{G-coherence}),                                             %d=n µÄ¸ßÎ¬Çé¿öÏÂ
\begin{equation}\label{16}
  G[\Phi(\ket{\psi}\bra{\psi})]=G(\ket{\psi}\bra{\psi})G[\Phi(\ket{\psi^+}\bra{\psi^+})].
\end{equation}
where $\ket{\psi^+}$ is the maximally coherent state (\ref{2}).

\textbf{Proof}. Obviously, when an arbitrary $d$-dimensional pure state $\ket{\psi}$ (\ref{1}) passes through a FSIO channel $\Phi$, by using Eq. (\ref{3}) and Eq. (\ref{5}), one can obtain that
\begin{equation}
  \Phi(\ket{\psi}\bra{\psi})=\sum_{n}{K_n S\ket{\psi^+}\bra{\psi^+}S^\dagger K_n^\dagger},
\end{equation}
with all Kraus operators $\{K_n\}$  satisfying Eq. (\ref{FSIO}).
So we can make use of the G-coherence to  quantify its full coherence,
\begin{eqnarray}
  G[\Phi(\ket{\psi}\bra{\psi})]&=&G(\sum_{n}{K_n S\ket{\psi^+}\bra{\psi^+}S^\dagger K_n^\dagger}) \nonumber\\
                               &=& G(\sum_{n}{U_{\pi}A_n S\ket{\psi^+}\bra{\psi^+}S^\dagger A_n^\dagger} U_{\pi}^\dag)  \nonumber\\
                               &=&  G(\sum_{n}{A_n S\ket{\psi^+}\bra{\psi^+}S^\dagger A_n^\dagger} ),
\end{eqnarray}
where we have used Eqs. (\ref{FSIO}) and (\ref{Uinvariant}).
Because $S$ and $A_n$ are both diagonal matrices under the reference basis,  $S A_n=A_n S$ holds and one can obtain
\begin{eqnarray}
  G[\Phi(\ket{\psi}\bra{\psi})]
  &=&G(S\sum_{n}{A_n\ket{\psi^+}\bra{\psi^+}A_n^\dagger} S^\dagger)\nonumber\\
  &=&G(SU_{\pi}^\dag U_{\pi}\sum_{n}{A_n\ket{\psi^+}\bra{\psi^+}A_n^\dagger}U_{\pi}^\dag U_{\pi}S^\dagger)\nonumber\\
  &=&G[U_{\pi}SU_{\pi}^\dag \Phi(\ket{\psi^+}\bra{\psi^+}) U_{\pi}S^\dagger U_{\pi}^\dag]\nonumber\\
  &=&G[S_{\pi}\Phi(\ket{\psi^+}\bra{\psi^+})S_{\pi}^\dagger],
\end{eqnarray}
where we have defined $S_{\pi}:=U_{\pi}SU_{\pi}^\dag$, which is still a diagonal matrix and can be written as $S_{\pi}=\sum_{i=1}^d a_i\sqrt{d}|\pi(i)\rangle\langle \pi(i)|=\sum_{i=1}^d \pi^{-1}(a_i)\sqrt{d}|i\rangle\langle i|$, $\{\pi^{-1}(a_i)\}$ is a permutation of $\{a_i\}$ which is inverse of $\pi$.

Let us denote the density matrix of $\Phi(\ket{\psi^+}\bra{\psi^+})$ as $M$, and denote $S_{\pi}MS_{\pi}^\dagger$ as $\rho^{out}$. Based on Eq. (\ref{G-coherence}), we can obtain
\begin{eqnarray}
  G[\Phi(\ket{\psi}\bra{\psi})]
  &=&d\prod_{i,j}^{i\neq j}{|\rho_{ij}^{out}|^{\frac{1}{d(d-1)}}}\nonumber\\
  &=&d\prod_{i,j}^{i\neq j}\bigg|\pi^{-1}(a_i)\sqrt{d}M_{ij}\pi^{-1}(a_j^*)\sqrt{d}\bigg|^{\frac{1}{d(d-1)}}  \nonumber\\
  &=&d^{2}\prod_{i,j}^{i\neq j}\bigg(|a_i a_j|^{\frac{1}{d(d-1)}}\cdot|M_{ij}|^{\frac{1}{d(d-1)}}\bigg)\nonumber\\
  &=&\bigg(d\prod_{i,j}^{i\neq j}|a_i a_j|^{\frac{1}{d(d-1)}}\bigg)\bigg(d\prod_{i,j}^{i\neq j}|M_{ij}|^{\frac{1}{d(d-1)}}\bigg)\nonumber\\
  &=& G(\ket{\psi}\bra{\psi})G[\Phi(\ket{\psi^+}\bra{\psi^+})], \label{24}
\end{eqnarray}
where we have used $\prod_{i,j}^{i\neq j}|\pi^{-1}(a_i) \pi^{-1}(a_j^*)|^{\frac{1}{d(d-1)}}=\prod_{i,j}^{i\neq j}|a_i a_j|^{\frac{1}{d(d-1)}}$.   Thus, the G-coherence evolution equation (\ref{16}) has been proved.  \hfill
$\square$

\textbf{Remark}. If $d=2$, i.e., we consider single-qubit systems, the G-coherence is equal to the $l_1$ norm of coherence. Thus, Eq. (\ref{16})  can be expressed as
\begin{equation}\label{l1}
 C_{l_1}[\Phi(\ket{\psi}\bra{\psi})]=C_{l_1}(\ket{\psi}\bra{\psi})C_{l_1}[\Phi(\ket{\psi^+}\bra{\psi^+})].
\end{equation}
Furthermore, by using the convex roof of G-coherence defined in Eq. (\ref{15}), one can also prove a coherence evolution equation under FSIO.

\textbf{Theorem 2}. If $\ket{\psi}$ is an arbitrary $d$-dimensional pure state (\ref{1}), after passing through a FSIO channel $\Phi$, its coherence satisfies the following evolution equation by using the convex roof of G-coherence (\ref{15}),                        %d=2 µÄ´¿Ì¬
\begin{equation}\label{18}
  \widetilde{G}[\Phi(\ket{\psi}\bra{\psi})]=\widetilde{G}(\ket{\psi}\bra{\psi})\widetilde{G}[\Phi(\ket{\psi^+}\bra{\psi^+})].
\end{equation}
where $\ket{\psi^+}$ is the maximally coherent state (\ref{2}).

\textbf{Proof. } If $\ket{\psi}$ goes through  a FSIO  channel $\Phi$, we can get,
\begin{eqnarray}
  \widetilde{G}[\Phi(\ket{\psi}\bra{\psi})]
  &=&\widetilde{G}(\sum_{n}{K_{n}S\ket{\psi^+}\bra{\psi^+}S^\dagger K_{n}^\dagger})\nonumber\\
  &=&  \widetilde{G}(\sum_{n}{U_{\pi}A_{n}S\ket{\psi^+}\bra{\psi^+}S^\dagger A_{n}^\dagger}U_{\pi}^\dagger)      \nonumber\\
    &=&  \widetilde{G}(\sum_{n}{A_{n}S\ket{\psi^+}\bra{\psi^+}S^\dagger A_{n}^\dagger}),\label{19}
\end{eqnarray}
where we have used $\widetilde{G}(U_{\pi}\rho U_{\pi}^\dag)=\widetilde{G}(\rho)$.
As $S$ and $A_{n}$ matrices are both diagonal matrices, $A_n S=S A_n$, we can get,
\begin{equation}
  \widetilde{G}(\sum_{n}{A_{n}S\ket{\psi^+}\bra{\psi^+}S^\dagger A_{n}^\dagger})
  =\widetilde{G}(S_{\pi}\Phi(\ket{\psi^+}\bra{\psi^+})S_{\pi}^\dagger),
\end{equation}
with  $S_{\pi}=U_{\pi}SU_{\pi}^\dag$.

Similar to the proof of Theorem 1, let us define $\Phi(\ket{\psi^+}\bra{\psi^+})$ as $\varrho$. According to the properties of the $\widetilde{G}$, suppose that we have already found an optimal decomposition for $\varrho$, $\varrho=\sum_{k^{'}}{p_{k^{'}}\ket{\psi_{k^{'}}}\bra{\psi_{k^{'}}}}$, to achieve the infimum of $\widetilde{G}(\varrho )$. It is worth noticing that this optimal decomposition can also achieve the infimum of $\widetilde{G}(S_{\pi}\varrho S_{\pi}^\dagger)$. Therefore,
\begin{equation}
  \widetilde{G}(S_{\pi}\varrho S_{\pi}^\dagger)
  ={\sum_{k^{'}}{p_{k^{'}}}G(S_{\pi}\ket{\psi_{k^{'}}}\bra{\psi_{k^{'}}}S_{\pi}^\dagger)}.
\end{equation}
For the $G(S_{\pi}\ket{\psi_{k^{'}}}\bra{\psi_{k^{'}}}S_{\pi}^\dagger)$, let's define $\ket{\psi_{k^{'}}}\bra{\psi_{k^{'}}}$ as $\varrho^{'}$. Similar to the derivation in Eq. (\ref{24}), we can get,
\begin{eqnarray}
  G(S_{\pi}\varrho^{'}S_{\pi}^\dagger)
  &=&d\prod_{ij}^{i\neq j}{|[S_{\pi}\varrho^{'} S_{\pi}^\dagger]_{ij}|^{\frac{1}{d(d-1)}}} \nonumber\\
  &=&d\prod_{ij}^{i\neq j}{|a_i\sqrt{d}\varrho^{'}_{_{ij}}a_j^*\sqrt{d}|^{\frac{1}{d(d-1)}}},\nonumber\\
    &=&d^{2}\prod_{ij}^{i\neq j}{|a_i\cdot a_j|^{\frac{1}{d(d-1)}}\cdot|\varrho_{ij}^{'}|^{\frac{1}{d(d-1)}}}\nonumber \\
  &=&G(\ket{\psi}\bra{\psi})G(\ket{\psi_{k^{'}}}\bra{\psi_{k^{'}}}).
\end{eqnarray}
Therefore, one has
\begin{eqnarray}
  \widetilde{G}(S_{\pi}\varrho S_{\pi}^\dagger)
  &=&{\sum_{k^{'}}{p_{k^{'}}}G(S_{\pi}\ket{\psi_{k^{'}}}\bra{\psi_{k^{'}}}S_{\pi}^\dagger)}\nonumber\\
  &=&\sum_{k^{'}}{p_{k{'}}}G(\ket{\psi}\bra{\psi})G(\ket{\psi_{k^{'}}}\bra{\psi_{k^{'}}})\nonumber\\
  &=&G(\ket{\psi}\bra{\psi})\sum_{k^{'}}{p_{k{'}}}G(\varrho^{'})\nonumber\\
  &=&\widetilde{G}(\ket{\psi}\bra{\psi})\widetilde{G}(\varrho)\nonumber\\
  &=&\widetilde{G}(\ket{\psi}\bra{\psi})\widetilde{G}(\Phi(\ket{\psi^+}\bra{\psi^+})).
\end{eqnarray}
Thus, Eq. (\ref{18}) holds. \hfill
$\square$

\subsection{Mixed single-qudit states}
Now we consider our results for mixed single-qudit states. The theorem of coherence evolution equation by using the G-coherence (\ref{G-coherence}) is as follows.

\textbf{Theorem 3.} If $\rho$ is a $d$-dimensional mixed state. We measure the coherence of $\rho$ with $G$, after it goes through a FSIO channel $\Phi$, we can get the following result,
\begin{equation}\label{th5}
{G}[\Phi(\rho)]={G}(\rho){G}[\Phi(\ket{\psi^+}\bra{\psi^+})].
\end{equation}
\textbf{Proof}. Let us just assume a series of $d$-dimensional Kraus operators $K_n$ of FSIO channel $\Phi$,
\begin{equation}
%\begin{split}
  K_n=U_{\pi}A_n=U_{\pi}\sum_i a_{ii}^{(n)}|i\rangle\langle i|=\sum_i a_{ii}^{(n)}|\pi(i)\rangle\langle i|,   %\begin{pmatrix} k^{(n)}_{1}& & &\\&k^{(n)}_{2} & & \\ & &k^{(n)}_{3} & \\& & &\cdots \\ \end{pmatrix}.
%\end{split}
\end{equation}
where $A_n=\mathrm{diag}\{\cdots,a_{ii}^{(n)},\cdots\}$ are  diagonal matrices under the reference basis satisfying $\sum_n A_n^\dag A_n=\idol$.
So the result of $G[\Phi(\ket{\psi}\bra{\psi})]$ could be that,
\begin{equation}\label{20}
\begin{split}
  G[\Phi(\rho)]
  &=G(\sum_{n}{K_n \rho K_n^\dagger })\\
  &=G[\sum_{n}\sum_{ij} a_{ii}^{(n)} \rho_{ij} a_{jj}^{(n)*} |\pi(i)\rangle\langle\pi(j)|]\\
  &=d\prod_{ij}^{i\neq j}{|\rho_{ij}(\sum_{n}{a_{ii}^{(n)}a_{jj}^{(n)*}})}|^{\frac{1}{d(d-1)}},\\
  &=\bigg(d\prod_{ij}^{i\neq j}{|\rho_{ij}}|^{\frac{1}{d(d-1)}}\bigg)\bigg(\prod_{ij}^{i\neq j}{|\sum_{n}{a_{ii}^{(n)}a_{jj}^{(n)*}}}|^{\frac{1}{d(d-1)}}\bigg).
\end{split}
\end{equation}
At the same time, we use the G-coherence to quantify $\Phi(\ket{\psi^+}\bra{\psi^+})$,
\begin{eqnarray}
%\begin{split}
  G[\Phi(\ket{\psi^+}\bra{\psi^+})]
  &=&G(\sum_{n}{K_n \ket{\psi^+}\bra{\psi^+} K_n^\dagger })\nonumber\\
  &=&\prod_{ij}^{i\neq j}{|\sum_{n}{a_{ii}^{(n)}a_{jj}^{(n)*}}}|^{\frac{1}{d(d-1)}},\label{21}
%\end{split}
\end{eqnarray}
and $G(\rho)$ is that,
\begin{equation}\label{22}
\begin{split}
  G(\rho)=d\prod_{ij}^{i\neq j}|\rho_{ij}|^{\frac{1}{d^2-d}}.\\
\end{split}
\end{equation}
From Eqs. (\ref{20}-\ref{22}), we can find that Eq. (\ref{th5}) holds. \hfill
$\square$

\textbf{Remark}. It is surprising that the equation (\ref{th5}) of coherence evolution equation for mixed states still holds, which is unlike entanglement  evolution equation for mixed states.

\begin{figure}
    \centering
    \includegraphics[scale=0.30]{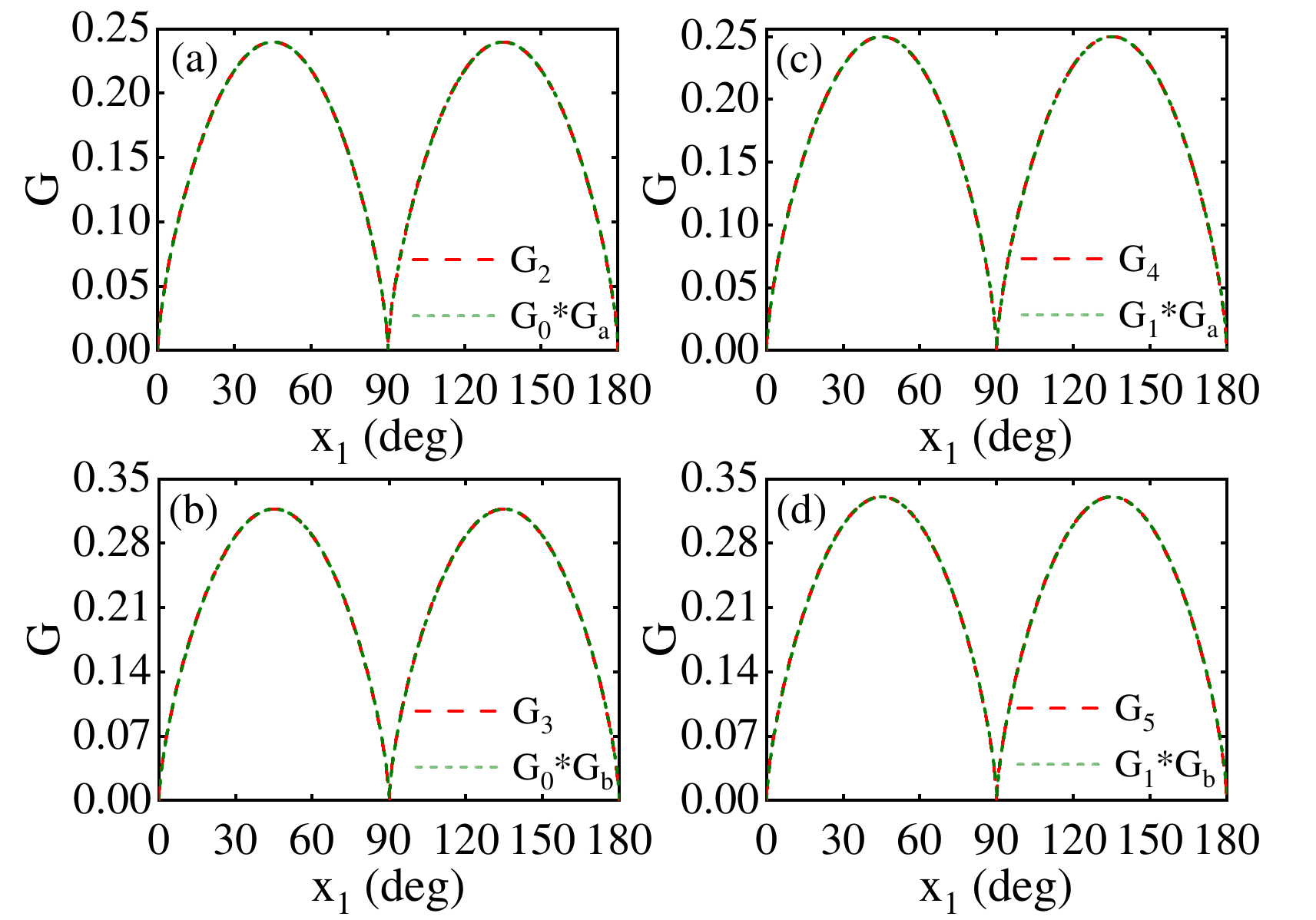}
    \caption{The verification for Eq.~(\ref{16}) in the pure qutrit states, including (a) $p_0=0.5$, $p_1=0.25$, $p_2=0.25$, $x_{0}=18^\circ$, (b) $p_0=0.5$, $p_1=0.125$, $p_2=0.375$, $x_{0}=18^\circ$, (c) $p_0=0.5$, $p_1=0.25$, $p_2=0.25$, $x_{0}=45^\circ$, and (d) $p_0=0.5$, $p_1=0.125$, $p_2=0.375$, $x_{0}=45^\circ$. The red dashed line and green dashed-dotted line
respectively represent the LHS and the RHS of the Eq.~(\ref{16})}\label{Fig2}
\end{figure}

Moreover, the theorem of coherence evolution equation for mixed states by using the convex roof of G-coherence (\ref{15}) is as follows.

\textbf{Theorem 4.} If $\rho$ is a mixed state in a $d$-dimensional system, after it goes through a FSIO channel $\Phi$, we can obtain the following inequality,
\begin{equation}\label{th6}
\begin{split}
\widetilde{G}[\Phi(\rho)]\leq \widetilde{G}(\rho)\widetilde{G}[\Phi(\ket{\psi^+}\bra{\psi^+})].
\end{split}
\end{equation}

\textbf{Proof.}
From Eq. (\ref{15}), suppose that we have already found an optimal decomposition for $\rho$ as $\rho=\sum_{k^{'}}{q_{k^{'}}\ket{\phi_{k^{'}}}\bra{\phi_{k^{'}}}}$, to achieve the infimum of $\widetilde{G}(\rho)$, i.e.,
\begin{eqnarray}
\widetilde{G}(\rho)&:=&\underset{\{p_k,\ket{\psi_k}\}}{\inf}\sum_{k}p_{k}G(\ket{\psi_{k}}\bra{\psi_{k}})\nonumber\\
&=&\sum_{k'}{q_{k'}G}(\ket{\phi_{k'}}\bra{\phi_{k'}})\nonumber\\
&=&\sum_{k'}{q_{k'}\widetilde{G}}(\ket{\phi_{k'}}\bra{\phi_{k'}}).
\end{eqnarray}
Therefore, by using Theorem 2 we can obtain
\begin{eqnarray}
&&\widetilde{G}(\rho)\widetilde{G}[\Phi(\ket{\psi^+}\bra{\psi^+})]\nonumber\\
&  =&\sum_{k'}{q_{k'}\widetilde{G}}(\ket{\phi_{k'}}\bra{\phi_{k'}})\widetilde{G}[\Phi(\ket{\psi^+}\bra{\psi^+})]\nonumber\\
&  =&\sum_{k'}{q_{k'}}\widetilde{G}[\Phi(\ket{\phi_{k'}}\bra{\phi_{k'}})].
\end{eqnarray}
In addition, we know $\widetilde{G}$ is a convex function. According to its properties, we could get:
\begin{eqnarray}
  \sum_{k'}{q_{k'}}\widetilde{G}[\Phi(\ket{\phi_{k'}}\bra{\phi_{k'}})]
  &\geq& \widetilde{G}[\sum_{k'}q_{k'}\Phi(\ket{\phi_{k'}}\bra{\phi_k'})]\nonumber\\
  &=& \widetilde{G}[\Phi(\rho)],\label{40}
\end{eqnarray}
where the last equality holds due to the linear property of $\Phi$.
Thus, inequality (\ref{th6}) holds.\hfill
$\square$

\textbf{Remark}. It is worth noticing that the left and right hands side of (\ref{th6}) may not be equal,  unlike Eq. (\ref{th5}) in Theorem 3. In inequality (\ref{40}), $\{q_{k^{'}},\ket{\phi_{k^{'}}}\}$ is the optimal decomposition for $\rho$ as $\rho=\sum_{k^{'}}q_{k^{'}}\ket{\phi_{k^{'}}}\bra{\phi_{k^{'}}}$, to achieve the infimum of $\widetilde{G}(\rho)$, but $\{q_{k^{'}},\Phi(\ket{\phi_{k^{'}}})\}$ is probably not the optimal decomposition for $\Phi(\rho)$ to achieve the infimum of $\widetilde{G}[\Phi(\rho)]$.

\section{EXAMPLES}
%We have added the verification result of Eq.~(\ref{16}) as shown in Fig.~\ref{Fig2}.

In the qutrit cases, we consider a  FSIO channel as follows
\begin{equation}\label{channel}
\begin{split}
\Phi(\rho)=\sum_{i=0}^{2}p_{i}K_{i}\rho K_{i}^\dag,
\end{split}
\end{equation}
where $K_{i}=xz^i$,
\begin{equation}\label{}
x=\begin{pmatrix}0&0&1\\1&0&0\\0&1&0\\
\end{pmatrix},\ \
z=\begin{pmatrix}1&0&0\\0&e^{i\frac{2\pi}{3}}&0\\0&0&e^{i\frac{4\pi}{3}}\\
\end{pmatrix},
\end{equation}
with $K_{i}^{\dag}K_{i}=\idol$ and $p_0+p_1+p_2=1$ ($p_0\neq p_1\neq p_2$).

In the pure qutrit states, we have chosen the initial state as,
\begin{equation}\label{psi1}
\begin{split}
|\psi_{1}\rangle=\frac{1}{\sqrt{1+\cos{x_0}^2}}(\sin{x_1}|0\rangle-\cos{x_1}|1\rangle+\cos{x_0}|2\rangle).
\end{split}
\end{equation}
Based on Eq.~(\ref{16}) we need to calculated the G-coherence of the $|\psi_{1}\rangle$ after passing througher the FSIO channel, the G-coherence of $|\psi_{1}\rangle$ and the G-coherence of the $|\psi^+\rangle$ after passing througher the FSIO channel. In the pure qutrit cases,
\begin{equation}\label{psi2}
\begin{split}
|\psi^+\rangle=\frac{1}{\sqrt{3}}(|0\rangle+|1\rangle+|2\rangle).
\end{split}
\end{equation}

As shown in Fig.~\ref{Fig2}, because we chose the different $p_{i}$, the corresponding $G[\Phi(|\psi^+\rangle\langle\psi^+|)]$ also corresponds to a different value. When $p_0=0.5$, $p_1=0.25$, $p_2=0.25$, the $|\psi^+\rangle$ after passing througher the FSIO channel is $G_{a}[\Phi(|\psi^+\rangle\langle\psi^+|)]=0.25$. When $p_0=0.5$, $p_1=0.125$, $p_2=0.375$, the $|\psi^+\rangle$ after passing througher the FSIO channel is $G_{b}[\Phi(|\psi^+\rangle\langle\psi^+|)]=\sqrt{7}/8$.

In the mixed qutrit states, we have chosen the usual initial state as,
\begin{equation}\label{psi3}
\rho=q|\psi^+\rangle\langle\psi^+|+\frac{(1-q)}{3}\idol,
\end{equation}
where $0 \textless q \textless 1$.

\begin{figure}
    \centering
    \includegraphics[scale=0.30]{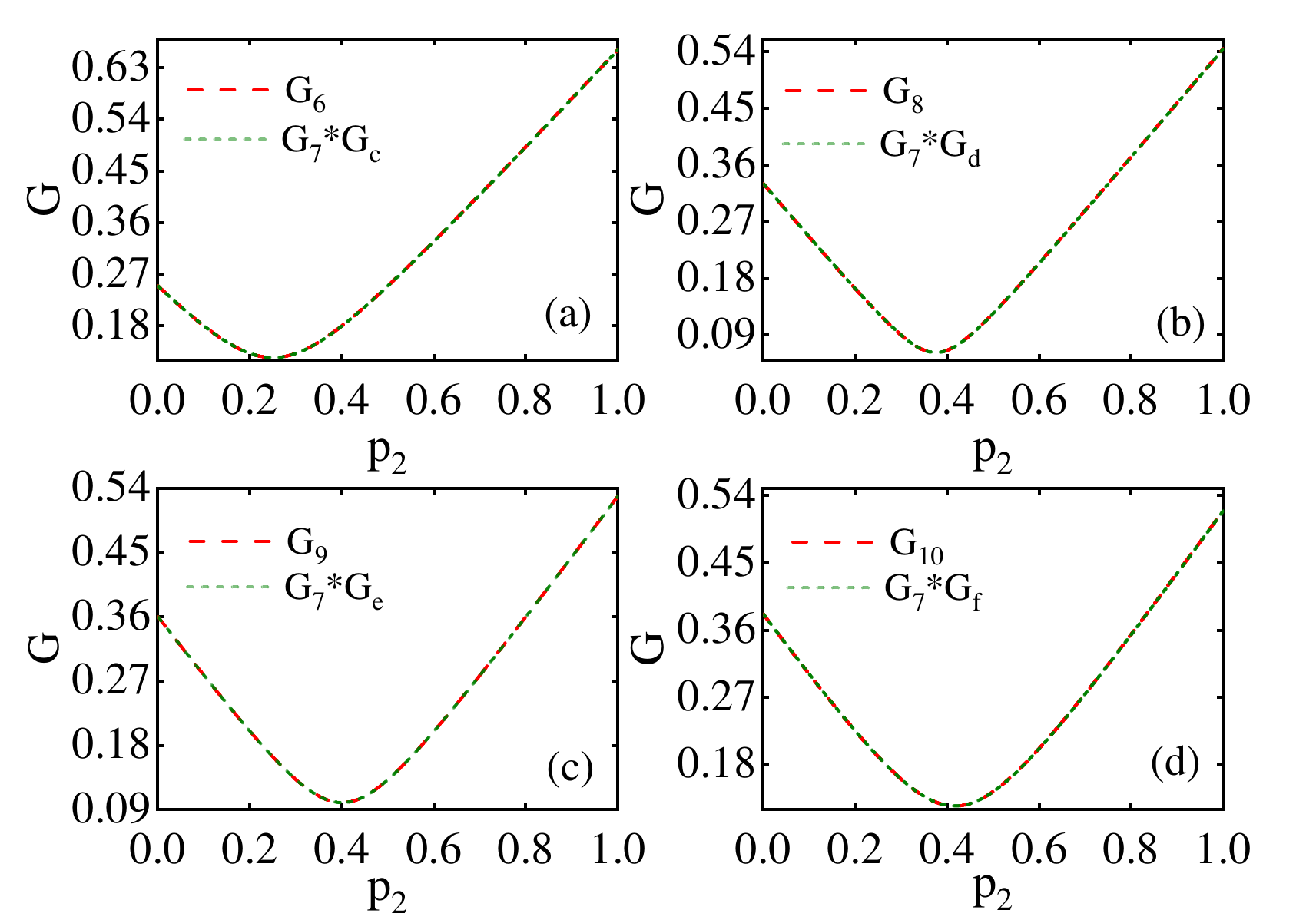}
    \caption{The verification for Eq.~(\ref{th5}) in the mixed qutrit states, we choose the different $p_i$ including (a)$p_0=0.5$, (b)$p_0=0.25$, (c)$p_0=0.2$ and (d)$p_0=1/6$. The red dashed line and green dashed-dotted line respectively represent the LHS and the RHS of the Eq.~(\ref{th5})}\label{Fig3}
\end{figure}

To verify the Eq.~(\ref{th5}), we need to calculated the G-coherence of the $\rho$ after passing througher the FSIO channel and  the G-coherence to quantify $\Phi(\ket{\psi^+}\bra{\psi^+})$ in different $p_i$. As shown in Fig.~\ref{Fig3}, the G-coherence of $\rho$ is $G_7(\rho)=0.5$.

\begin{figure}
    \centering
    \includegraphics[scale=0.30]{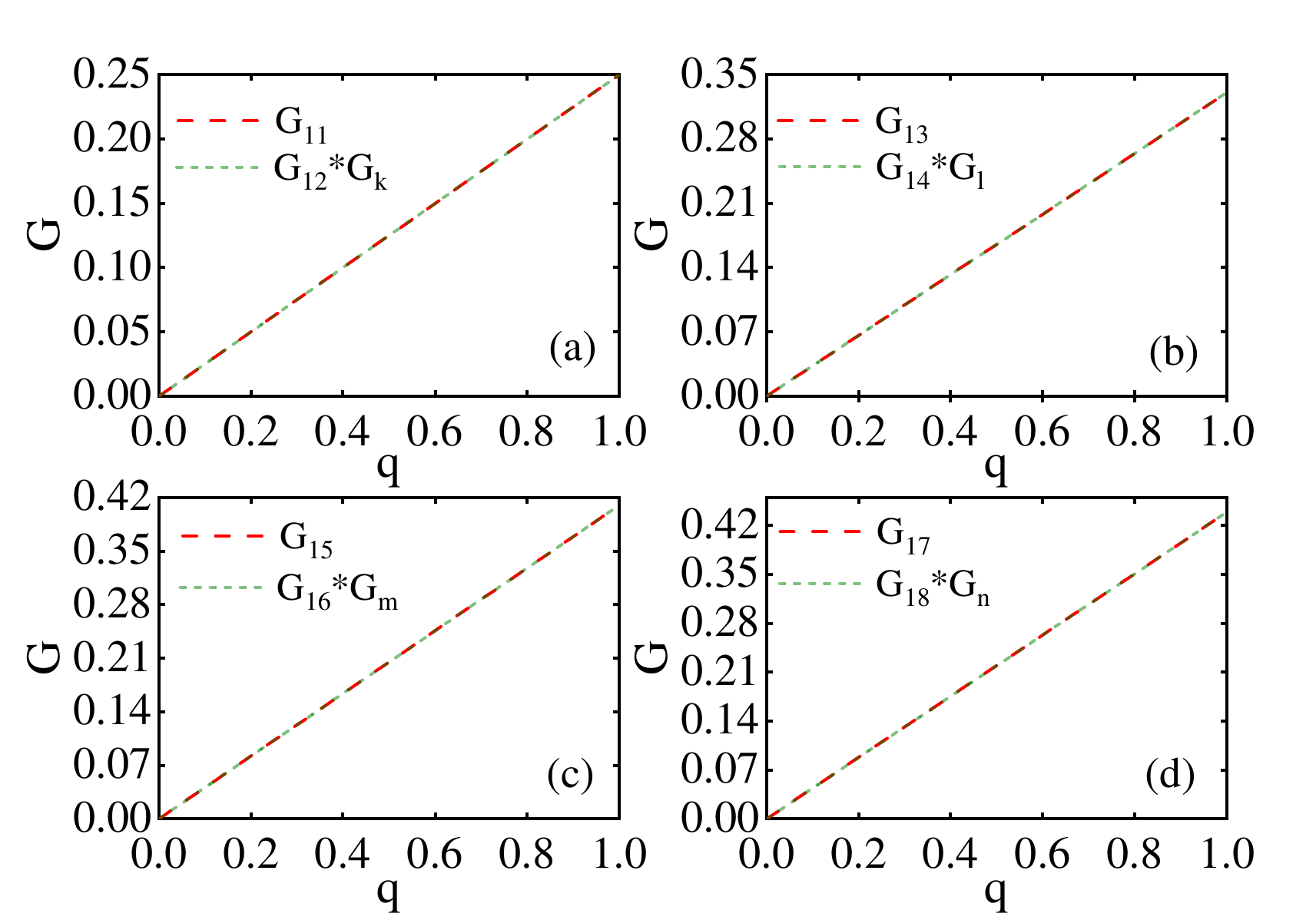}
    \caption{The verification for Eq.~(\ref{th5}) in the mixed qutrit states, we choose the different $q$ including (a) $p_0=0.5$, $p_1=0.25$, $p_2=0.25$, (b) $p_0=0.5$, $p_1=0.125$, $p_2=0.375$, (c) $p_0=0.5$, $p_1=0.0625$, $p_2=0.4375$, and (d) $p_0=0.5$, $p_1=1/24$, $p_2=11/24$. The red dashed line and green dashed-dotted line respectively represent the LHS and the RHS of the Eq.~(\ref{th5})}\label{Fig4}
\end{figure}

As shown in Fig.~\ref{Fig4}, it was also noted that at the determined $p_i$, different $q$ values were adjusted to obtain different mixed states $\rho$, in order to verify  the Eq.~(\ref{th5}).

\section{Discussions and conclusions}

In addition to FSIO, there are some other special channels that follow the coherence evolution equation. For example, let us consider a generalized amplitude damping channel,
\begin{eqnarray}
  &&\widetilde{K}_0=\sqrt{p}\begin{pmatrix}1&0\\0&\sqrt{1-\epsilon} \\
\end{pmatrix},\ \
\widetilde{K}_1=\sqrt{p}\begin{pmatrix}0&\sqrt{\epsilon}\\0&0 \\
\end{pmatrix},\nonumber\\
&&\widetilde{K}_2=\sqrt{1-p}\begin{pmatrix}\sqrt{1-\epsilon}&0\\0&1 \\
\end{pmatrix},\ \
\widetilde{K}_3=\sqrt{1-p}\begin{pmatrix}0&0\\\sqrt{\epsilon}&0 \\
\end{pmatrix}.\nonumber
\end{eqnarray}
When $p=1$, it reduces to a standard amplitude damping channel, which has been discussed in Ref. \cite{exp}. One can check that under the generalized amplitude damping channel $\{\widetilde{K}_i\}$ Theorems 3-6 still hold. The reason is that both $\widetilde{K}_1 \rho \widetilde{K}_1^\dag$ and $\widetilde{K}_3\rho \widetilde{K}_3^\dag$ have zero off-diagonal elements. Thus, under FSIO Kraus operators (\ref{FSIO}) with $K\rho K^\dag$ zero off-diagonal element operators like  $\widetilde{K}_1$ and  $\widetilde{K}_3$, the evolution equation for quantum coherence (Theorems 3-6) still holds.

In conclusion, we have proved the coherence evolution equation, which is applicable to the G-coherence and convex roof of G-coherence after quantum states passing through the FSIO channels, including pure states and mixed states. Particularly, for a mixed state, G-coherence still satisfies the coherence evolution equation under the condition of FSIO. Our theorems provide a novel method to quantify the coherence of quantum states after passing through FSIO channels and would play an important role in the simplification of dynamical coherence measure.

\section*{ACKNOWLEDGMENTS}
This work is supported by the National Natural Science Foundation of China (Grant No. 11734015), and K.C. Wong Magna Fund in Ningbo University.

\end{document}